\documentclass[prl,aps,twocolumn]{revtex4}
\usepackage{amsmath,amssymb,graphicx}

\usepackage{amsmath,amssymb,graphicx}
\usepackage{pxfonts}
\usepackage{graphicx}

\newcommand{\unit}[1]{\,\mathrm{#1}} 

\begin{document}

\title{Two-Dimensional Electrostatic Lattices for Indirect Excitons}

\author{M. Remeika, M.M. Fogler, and L.V. Butov}
\affiliation{Department of Physics, University of California at San
Diego, La Jolla, CA 92093-0319}

\author{M. Hanson and A.C. Gossard}
\affiliation{Materials Department, University of California at Santa
Barbara, Santa Barbara, California 93106-5050}

\begin{abstract}
We report on a method for the realization of two-dimensional electrostatic lattices for excitons using patterned interdigitated electrodes. Lattice structure is set by the electrode pattern and depth of the lattice potential is controlled by applied voltages. We demonstrate square, hexagonal, and honeycomb lattices created by this method.
\end{abstract}


\date{\today}

\maketitle

Studies of particles in periodic potentials are fundamental to condensed matter physics. While originally experimental studies concerned electrons in crystal lattices, a variety of systems with particles in artificial lattice potentials are actively investigated at present. Controlling the parameters of an artificial lattice provides a tool for studying the properties of particles confined to the lattice and, to some extent, for emulating condensed matter systems. Cold atoms in an optical lattice present a prominent example of particles in artificial lattices. Phenomena originally considered in context of condensed matter systems, such as the Mott insulator -- superfluid transition, can be studied in the system of cold atoms in optical lattices \cite{Greiner02}.

\begin{figure}
\begin{center}
\includegraphics[width=7.5cm]{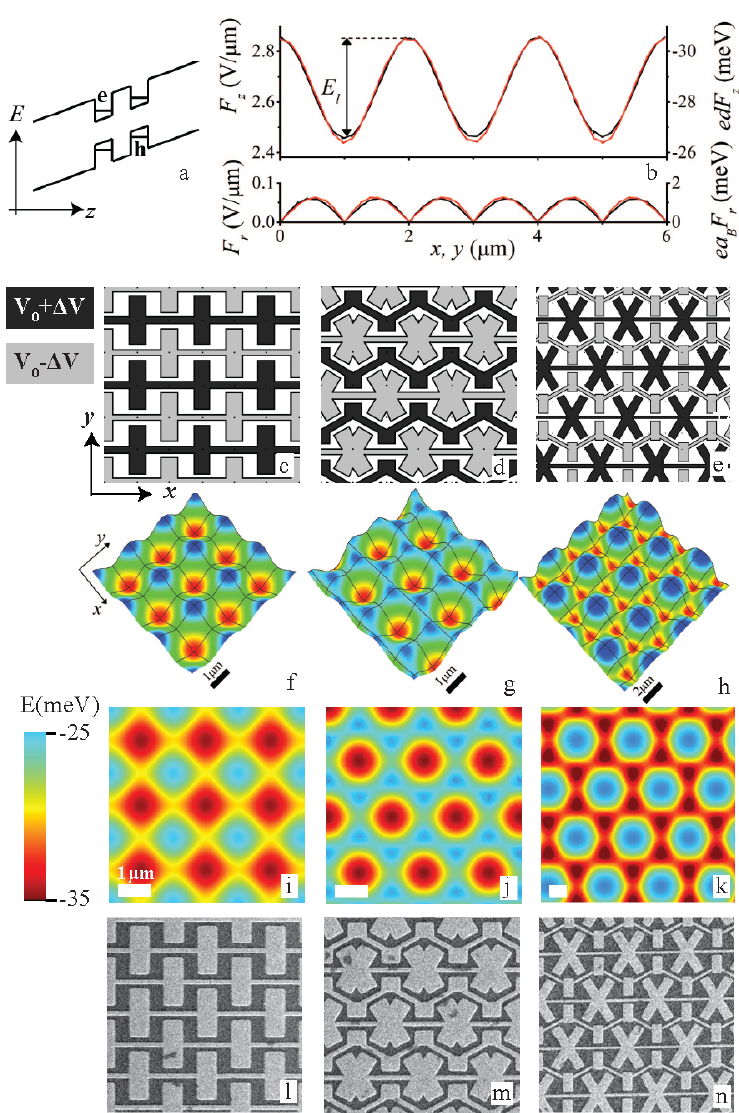}
\caption{(color online) (a) Energy diagram of the CQW. (b) Simulated electric field $F_z$ and exciton energy $e d F_z$ along $x$ (black) and $y$ (red) for square lattice. Lower plot shows lateral electric field $F_r = (F_x^2 + F_y^2)^{1/2}$ and $e F_r a_B$. (c), (d), (e) Electrode schematics for square, triangular, and honeycomb lattices, respectively. (f-k) Simulated exciton energy for these electrode patterns. $\Delta V = 1\unit{V}$ (see Supplementary information). (l-n) SEM images of the electrode patterns.
\label{fig:schematic}}
\end{center}
\end{figure}

Excitons in artificial lattices present a condensed matter system of particles in periodic potentials \cite{Zimmermann98, Hammack06, Rudolph07, Lai07, Remeika09, Cerda10, Byrnes10, Flayac11, Winbow11}. In particular, artificial periodic potentials, both static and moving, can be created for indirect excitons \cite{Zimmermann98, Hammack06, Remeika09, Winbow11}. An indirect exciton in coupled quantum wells (CQW) is a bound state of an electron and a hole in separate QWs (Fig. \ref{fig:schematic}a). Due to a dipole moment of indirect excitons $e d$ ($d$ is close to the distance between the QW centers), potential landscapes for excitons $E(x,y) = e d F_z(x,y) \propto V(x,y)$ can be created using a laterally modulated gate voltage $V(x,y)$ ($F_z$ is the $z$-component of electric field in the CQW layers) \cite{Zimmermann98, Hammack06, Remeika09, Winbow11, Hagn95, Huber98, Gartner06, Chen06, High07, High08, Grosso09, High09}. Furthermore, due to their long lifetimes, orders of magnitude longer than that of regular excitons, indirect excitons can travel in electrostatically created potentials over large distances before recombination \cite{Remeika09, Winbow11, Hagn95, Gartner06, High07, High08, Grosso09, High09}. Also, due to their long lifetimes, these bosonic particles can cool to temperatures well below the temperature of quantum degeneracy \cite{Butov01}. Therefore, the system of indirect excitons in electrostatic lattices gives an opportunity to study transport of cold bosons in periodic potentials.

Linear lattice potentials with energy modulation in one dimension were created for indirect excitons by interdigitated gates \cite{Zimmermann98, Remeika09, Winbow11}. However, a number of phenomena, including the Mott insulator -- superfluid transition, require in-plane energy modulation in both dimensions \cite{Greiner02}. A two-dimensional (2D) lattice for excitons can be generated by a single electrode with a periodic array of holes \cite{Hammack06}. The lateral modulation of $F_z$, which determines the lattice depth, can be controlled by changing the voltage applied to the electrode. However, within this method, changing the lattice amplitude is accompanied by changing the average electric field $F_z^{avg}$ and, in turn, lifetime and density of indirect excitons. An independent control $F_z^{avg}$ and the lateral modulation of $F_z$ can be realized using multiple electrodes separated by insulating layer(s) \cite{Krauss04}. Superposition of the fields from such electrodes can create the desired pattern of $F_z(x,y)$. However, within this method, the semiconductor structure is in series with the deposited insulator layer(s). Therefore, considerable fraction of the applied voltage, determined by the ratio between the conductance of the insulator and semiconductor layers, can drop in the deposited insulator. Also, this fraction and, in turn, $F_z(x,y)$ may depend on the optical excitation.

Here, we present a method for creating 2D electrostatic lattices for indirect excitons that explores the opportunity to control exciton energy by electrode density \cite{Kuznetsova10}. We demonstrate that 2D lattices for excitons can be produced by patterned interdigitated gates. The lattice constant and lattice structure are determined by the electrode pattern. Figures~\ref{fig:schematic}c, d, and e show the electrode patterns for creating square, triangular, and honeycomb lattices, respectively. The corresponding simulated exciton potential profiles are shown in Fig.~\ref{fig:schematic}f-k. The average field $F_z^{avg}$ and spatial modulation of $F_z$ can be independently controlled by voltages $V_0$ and $\Delta V$. $F_z^{avg}$ realizes the indirect exciton regime and controls the exciton lifetime. Modulation of $F_z$ forms the lattice potential (Fig.~\ref{fig:schematic}b). The lattice amplitude can be controlled in situ by $\Delta V$. The in-plane electric field in the lattice $F_{xy}$ is small so that it does not cause the exciton dissociation: $e F_{xy} a_B \ll E_{ex}$, $a_B \sim 20$ nm and $E_{ex} \sim 4$ meV are the Bohr radius and binding energy for the indirect excitons, respectively \cite{Dignam91,Szymanska03} (Fig.~\ref{fig:schematic}b).

Advantages of this method include: (i) A variety of 2D lattice structures for excitons can be realized; (ii) The depth of the lattice potential can be controlled in situ by voltage; (iii) The average field can be controlled by voltage independently from lattice depth; (iv) Smooth 2D lattice potentials are realized by the electrode patterns; (v) The lattice device can be fabricated using single layer lithography with no deposited insulator layer.

We demonstrate experimental proof of principle for creating 2D lattices for excitons by this method. A square lattice potential (Fig. \ref{fig:schematic}i) was used for the demonstration. CQW structure was grown by MBE. $n^+$-GaAs layer with $n_{Si} = 10^{18}\unit{cm}^{-3}$ serves as a homogeneous bottom electrode. Semitransparent top patterned electrodes were fabricated by evaporating $2\unit{nm}$ Ti and $7\unit{nm}$ Pt. CQW with $8\unit{nm}$ GaAs QWs separated by a $4\unit{nm}$ Al$_{0.33}$Ga$_{0.67}$As barrier were positioned 100 nm above the $n^+$-GaAs layer within an undoped $1\unit{\mu m}$ thick Al$_{0.33}$Ga$_{0.67}$As layer. Excitons were photogenerated by a Ti:Sapphire laser tuned to the energy of direct excitons in this sample ($\approx 786$~nm). Exciton density was controlled by the laser excitation power $P_{ex}$. Photoluminescence (PL) images of the exciton cloud were captured by a CCD with a filter selecting photon wavelengths $\lambda = 800 \pm 5\unit{nm}$ covering the spectral range of the indirect excitons. The spectra were measured using a spectrometer with resolution $0.18\unit{meV}$. Experiments were done at $T_{\text{bath}} = 1.6\unit{K}$.

\begin{figure}
\begin{center}
\includegraphics[width=7.5cm]{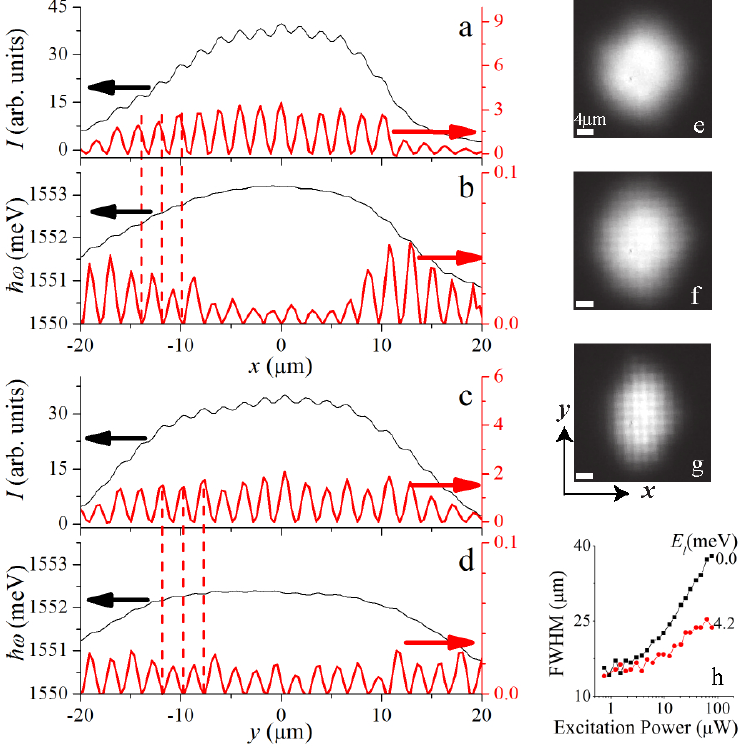}
\caption{(color online) (a) PL intensity and (b) energy of excitons in a square lattice along $x$ (black). The same data with a smooth curve subtracted (red). (c, d) Similar data along $y$. Dashed lines are guides to the eye. Laser spot FWHM is $17\unit{\mu m}$ along $x$ and $14\unit{\mu m}$ along $y$. $P_{ex}=40 \mu$W, $E_l=4.2$ meV. (e, f, g) Exciton PL images in a square lattice for $P_{ex}=11\unit{\mu W}$ at $E_l =$ 0, 2.1, and 4.2 meV respectively. (g) FWHM of exciton cloud emission along $x$ for $E_l = 0$ (black) and 4.2 meV (red) vs $P_{ex}$.
\label{fig:transport_modulation}}
\end{center}
\end{figure}

Figure \ref{fig:transport_modulation}a-d shows the emission profiles for excitons in a square lattice along $x$ or $y$. Each point in the $x$-profiles was obtained by averaging over 5 lattice sites along $y$ and \textit{vice versa} to reduce the noise in the data. Another source of averaging is finite optical resolution, see below (note that averaging reduced the amplitude of the spatial modulations discussed below). The quantity $\hbar \omega$ in Fig.~\ref{fig:transport_modulation}b and d stands for the spectral average $\hbar \omega = M_1 / I$, where $I = \int I_{\omega^\prime} d \omega^\prime$ is the total PL intensity and $M_1 = \int I_{\omega^\prime} \hbar \omega^\prime d \omega^\prime$ is its first spectral moment. As one can see in Fig.~\ref{fig:transport_modulation}, both $I$ and $\hbar \omega$ are modulated with the period matching the lattice constant revealing the exciton confinement in the 2D lattice. The intensity maxima match the energy minima demonstrating exciton collection in the lattice sites.

We also probed exciton transport in the lattice. Figure \ref{fig:transport_modulation}e-g shows spatial images of exciton PL at three different lattice depths. As the lattice depth is turned up the exciton cloud width becomes smaller and locations of the lattice sites become apparent in the PL image. Figure \ref{fig:transport_modulation}h shows the full width at half maximum (FWHM) of the exciton cloud PL as a function of $P_{ex}$ for lattice depths $E_l = 0$ and 4.2 meV. At low exciton densities, the emission spot is essentially equal in size to the laser excitation spot indicating that excitons are localized and do not travel outside the laser excitation spot. At high exciton densities, the emission spot is larger than the laser excitation spot indicating that excitons are delocalized and travel outside the laser excitation spot (Fig. \ref{fig:transport_modulation}h). In similarity to the localization-delocalization transition studied in linear lattices \cite{Remeika09}, this behavior corresponds to the exciton localization in the combined lattice potential and disorder potential at low densities and exciton delocalization due to screening of the potential by the repulsively interacting excitons (the amplitude of the disorder potential in the sample is $\sim 0.6$ meV). Figure \ref{fig:transport_modulation}h shows that a higher exciton density and, in turn, higher interaction energy is required for screening the potential with a higher lattice amplitude, in agreement with this model.

\begin{figure}
\begin{center}
\includegraphics[width=7.5cm]{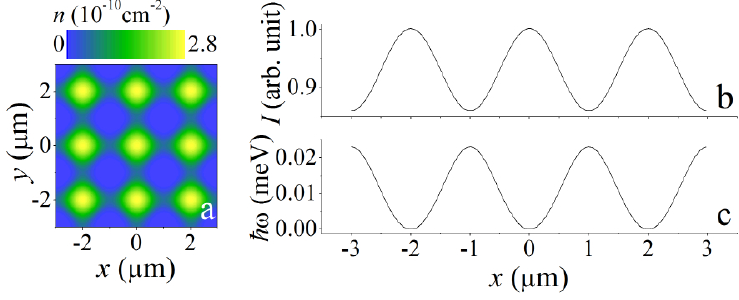}
\caption{(color online) Simulations for a square lattice. (a) Exciton density. Emission (b) intensity and (c) energy with averaging in $y$ and accounting for spatial resolution as in the experiment. $E_l=4.2$ meV, $\gamma = 7$, $T = 4$ K, $\mathrm{NA} = 0.245$.}
\label{fig:theory}
\end{center}
\end{figure}

In order to examine this agreement quantitatively, we considered a mean-field model~\cite{Remeika09} where the local density $n(\mathbf{r})$ of bright excitons is the solution of the equation
\begin{equation}\label{eqn:n}
\varepsilon(n) \equiv T \ln\left(1 - e^{n\, /\, 2\nu_1 T}\right)
 = E(\mathbf{r}) + \frac{\gamma}{\nu_1}\, n - \zeta\,.
\end{equation}
Here $\nu_1 = m / (2\pi \hbar^2)$ is the density of states per spin, $\gamma$ is the dimensionless interaction constant, $\zeta$ is the exciton electrochemical potential, and $T$ is the exciton temperature. Within this model, the first moment of the exciton emission energy proves to be
\begin{equation}
M_1 = (\zeta - \varepsilon) n + 2 \nu_1\, T^2\, \text{Li}_2\left(
e^{\varepsilon / T} \right)\,,
\label{eqn:E}
\end{equation}
where $\text{Li}_2(z)$ is the dilogarithm function. From these two quantities the local PL intensity and energy can be calculated (see above). For a more accurate comparison with the experiment, we also included the effect of the finite spatial resolution of our optical system. The choice of the fitting parameters in the model $\delta \sim 1\, \mu\text{m}$ for the defocussing parameter (see Supplementary information), $T = 3.6\unit{K}$, $\zeta = 5.0\unit{meV}$, and $\gamma = 2.3$ leads to a reasonable agreement between the simulations (Fig.~\ref{fig:theory}) and the experiment (Fig.~\ref{fig:transport_modulation}). However, this fitting should not be overemphasized because of a number of the fitting parameters and approximations made in the model.

In conclusion, we present a method for producing 2D lattices for indirect excitons and experimental proof of principle for this method. This work was supported by the DOE Office of Basic Energy Sciences under award DE-FG02-07ER46449. MF is supported by the UCOP.


\section{Supplementary Information}

\subsection{Electrostatic simulations}

The electrostatic potential $\phi(\mathbf{r})$ in the system in the absence of excitons was calculated numerically using COMSOL Multiphysics 4.0 software package. The system was modeled as a rectangular box $1\unit{\mu m}$ thick in the $z$ direction and five or more lattice periods wide in the $x$ and $y$ directions with the electrode pattern embedded into the top surface of the box. The potential was calculated by solving the Laplace equation in the volume of the box. At the electrode surfaces the boundary condition of constant potential was imposed, e.g., at the ground plane (bottom surface) we have $\phi = 0$. At all the other surfaces of the simulation box the condition of vanishing electric displacement, $D_\perp = 0$, was chosen.  The $z$-component of the electric field at $100\unit{nm}$ from the bottom plane (corresponding to the location of the quantum wells) was used to calculate exciton energy $E = e d F_z$.

\subsection{Optical resolution effects}

The spatial resolution of the optical system is described by its point spread function (PSF) $P(\mathbf{r})$. We used the following common model~\cite{Hopkins55} for the PSF
\begin{equation}
P(\mathbf{r}) = \left|\, \int d^2 q\, \Theta(Q - |\mathbf{q}|\,)\,
e^{i \mathbf{q} \mathbf{r} - i \delta^2 q^2 /2}\, \right|^2.
\label{eqn:P}
\end{equation}
This PSF has a finite width determined by the length scale
$Q^{-1} \equiv \lambda / (2 \pi \text{NA}) = 0.46\,\mu\text{m}$
set by the numerical aperture $\text{NA}$ of the system and by another length scale $\delta \sim 1\, \mu\text{m}$ that describes defocussing. The observable intensity $I(\mathbf{r})$ and its first spectral moment $M_1(\mathbf{r})$ were calculated by taking the convolution of the PSF and the ``ideal'' $I$ and $M_1$ derived from the mean-field theory
described in the main text.


\end{document}